\documentclass{article}
\usepackage{RR}
\usepackage{amssymb}
\usepackage{rotating}

\newtheorem{theorem}{Theorem}
\newtheorem{remark}{Remark}
\newtheorem{lemma}{Lemma}
\newtheorem{corollary}{Corollary}
\newtheorem{proposition}{Proposition}
\newtheorem{definition}{Definition}
\newenvironment{proof}{\textbf{Proof:}}{\hfill $\square$\bigskip\\}

\usepackage{hyperref}

\URRocq
\RRtitle{Complexité de résolution d'un système parametré d'\'equations et d'inéquations polynomiales}
\RRetitle{Complexity of Resolution of Parametric Systems of Polynomial Equations and Inequations \hspace{0.1cm}}
\RRauthor{Guillaume Moroz}
\RRtheme{\THSym}
\RRprojet{Salsa}
\RRdate{Mai 2006}
\RRresume{
On considère un système de $n$ équations polynomiales et $r$ inéquations en $n$ inconnues et $s$ paramètres. Le degré des polynômes considérés est majoré par $d$ et leurs coefficients sont rationels de taille binaire au plus $\sigma$. D'un point de vue réel, résoudre un tel système revient souvent à décrire un semi-algébrique de l'espace des paramètres au-dessus duquel le nombre de solutions réels du système parametré considéré est constant. D'après les travaux de Lazard et Rouillier, on peut obtenir ce semi-algébrique par le calcul d'une \emph{variété discriminante}.
Dans ce rapport, nous nous restreignons au cas où le systèmes d'équations donnés en entrée est zéro-dimensionel pour une spécialisation générique des paramètres, ce qui correspond à une situation courante dans les applications. Dans ce cas, nous proposons une méthode \emph{déterministe} pour calculer la variété discriminante \emph{minimale} en réduisant le problème à un problème d'élimination. De plus, nous prouvons que le degré de la variété discriminante minimale est majorée par $D:=(n+r)d^{(n+1)}$ et que la complexité de notre méthode est de $\sigma^{\mathcal{O}(1)} D^{\mathcal{O}(n+s)}$ opérations binaires sur une machine de Turing.
}
\RRmotcle{Système polynomial parametré, Variété discriminante, \'Elimination, Complexité déterministe}
\RRabstract{
Consider a system of $n$ polynomial equations and $r$ polynomial inequations in $n$ indeterminates of degree bounded by $d$ with coefficients in a polynomial ring of $s$ parameters with rational coefficients of bit-size at most $\sigma$. From the real viewpoint, solving such a system often means describing some semi-algebraic sets in the parameter space over which the number of real solutions of
the considered parametric system is constant. Following the works of Lazard and
Rouillier, this can be done by the computation of a {\em discriminant variety}.
In this report we focus on the case where for a generic specialization of the parameters the system of equations generates a radical zero-dimensional ideal, which is usual in the applications. In this case, we provide a {\em deterministic} method computing the {\em minimal} discriminant variety reducing the problem to a problem of elimination. Moreover, we prove that the degree of the computed minimal discriminant variety is bounded by $D:=(n+r)d^{(n+1)}$ and that
the complexity of our method is $\sigma^{\mathcal{O}(1)} D^{\mathcal{O}(n+s)}$ bit-operations on a deterministic Turing machine.
}
\RRkeyword{Parametric polynomial system, Discriminant variety, Elimination, Deterministic complexity}

\begin{document}
\makeRR
\section{Introduction}
The parametric polynomial systems are used in many different fields such as robotics, optimization, geometry problems, and so on.
In \cite{rouillier/lazard} the authors introduce the notion of discriminant variety which allows them to
split the parameter space in open cells where the number of real solutions is constant
. Even if it is efficient in a practical point of view, their algorithm is based Gr\"obner bases computations, whose complexity is not yet well understood. Thus it does not allow us to give a better bound than the worst case's one, which is in exponential space (\cite{conf/eurosam/Giusti84}).

In this article we prove that, under some assumptions, the computation of the minimal discriminant variety of a 
parametric system is reducible to the FPSPACE problem of general elimination \cite{Matera97}. The proof of the reduction correctness presented here is non trivial.  The reduction itself is simple and preserves the sparsity of the input system.\\

Our input is a system of polynomial equations and inequations of degrees bounded $d$, which can be written as:
$$\left\{ \begin{array}{c}f_1(t,x)=0\\\vdots\\f_n(t,x)=0\end{array}\right.
  \mbox{ and }
  \left\{ \begin{array}{c} g_1(t,x)\neq0\\\vdots\\g_r(t,x)\neq0\end{array}\right.\quad(t,x)\in\mathbb{C}^s\times\mathbb{C}^n$$
where $x$ are the unknowns and $t$ are the parameters. Moreover, for all specializations in an open ball of the parameters space, the system has a finite number of simple solutions in the unknowns. Such a system will be said \emph{generically simple} (see Definition \ref{gensimp}).\\
We prove that the degree of the minimal discriminant variety of a \emph{generically simple} parametric system is bounded by
$$(n+r)d^{n+1}$$
Our algorithm for \emph{generically simple} parametric systems runs in
$$\sigma^{\mathcal{O}(1)}(n+r)^{\mathcal{O}(n+s)}d^{\mathcal{O}(n(n+s))}$$
bit-operations on a deterministic Turing machine.\\

When we aim to solve a parametric system, we face two kinds of issues: either we want to describe the solutions in terms of the parameters, or else we want to classify the parameters according to properties of the parametric system's solutions.
Different methods have been developped to treat these two problems.

Regarding the first one, many algorithms exist in the literature. Among them we may cite rational parametrizations \cite{MR1959170}, 
triangular sets decompositions \cite{MR1826878}, 
comprehensive Gr\"obner bases \cite{Weispfenning92,conf/issac/GrigorievV00}. 
We may also mention numerical algorithms such as the Newton-Raphson or the homotopy continuation method \cite{conf/issac/Verschelde04,MR2161992}, which can be used after a specialization of the parameters.

Regarding the second problem on the parameters classification, few algorithm are available, whereas many applications face it,
such as parametric optimization (\cite{FRPM:ijs:2006}), robot modelling (\cite{conf/adg/CorvezR02}), geometry problems (\cite{Yang:2005:OPM}) or control theory (\cite{conf/issac/AnaiHY05}) for example. The C.A.D. \cite{Collins75,conf/issac/BrownM05} is the most widespread method. It computes an exhaustive classification, leading to a complexity doubly exponential in the number of unknowns. Some of the algorithms mentioned above (\cite{MR1826878, conf/issac/GrigorievV00}) may also return such kinds of classifications. Especially in \cite{conf/issac/GrigorievV00} the authors compute a complete partition of the parameters space in constructible sets where the vector of multiplicities of the system's solutions is constant. The time complexity of their algorithm is $d^{\mathcal{O}(n^2s)}$. However, they don't consider inequations and their algorithm is not meant to be implemented. The minimal discriminant variety is included in both of the precedent computations. It describes the maximal open subset of the parameters space where the system's solutions evolve regularly. The computation of this variety is indeed sufficient for a lot of applications.\\

Our method is a reduction to the general elimination problem. The elimination problem has been widely analysed in the past decades, as it is a key step for quantifier elimination theory (in \cite{TCS::Heintz1983,MR1624458,BasPolRoy96,Basu99a} for example), computation of the dimension of an ideal (\cite{oai:CiteSeerPSU:552837} among others) or implicitization theory (see \cite{Cox92}). Different techniques and software have been developed. We may mention sparse resultants (see \cite{Cox-Little-O_Shea/98} and references therein), linear system reductions (in \cite{oai:CiteSeerPSU:552837} for example), linear systems parametrized with straight-line programs (see \cite{MR1624458,MR1414452}), parametric geometric resolution (\cite{GiustiSchost:1999,MR1959170}) or Gr\"obner bases (see \cite{Buchberger:1976:TBR} and \cite{Faugere:2002:NEA,MR2093185} for the last improvements).\\

This article is divided in three parts. In the first one we reduce the problem of computing the minimal discriminant variety to the elimination problem. In the second part, we bound the degree of the minimal discriminant variety. And in the last part we give some examples.
\subsection*{Definition and notation}
\label{base}
In the following, we assume that
$$f_1,\cdots,f_n,g_1,\cdots,g_r\in\mathbb{Q}[T_1,\cdots,T_s][X_1,\cdots,X_n]$$
are some polynomials in degrees $d_i=deg(f_i)$ and $d'_j=deg(g_j)$ for $1\leq i \leq n$ and $1\leq j \leq r$.
We denote by $\mathbb{P}_n$ the projective closure of $\mathbb{C}^n$ and by $\pi:\mathbb{C}^s\times\mathbb{C}^n\rightarrow \mathbb{C}^s$ (resp. $\overline{\pi}:\mathbb{C}^s\times\mathbb{P}_n\rightarrow \mathbb{C}^s$) the canonical projection onto the parameters space.
The exponent $h$ (resp. $h_i$) of a polynomial or of an ideal denotes its homogenization by the variable $X_0$ with respect to the variables $X_1,\cdots,X_n$ (resp. its homogenization by the variable $X_i$ with respect to the variables $X_0,\cdots,\hat{X_i},\cdots,X_n$) . The term \emph{parameters} will refer to the variables $T_1,\cdots,T_s$, while the term \emph{unknowns} will refer to the variables $X_1,\cdots,X_n$.

Finally we use the following notation for the specialization of some variable. For $I\subset\mathbb{Q}[Y_1,\cdots,Y_k,Z]$ and $a\in\mathbb{Q}$, we denote:
$$I_{|Z=a}:=(I+\left<Z-a\right>)\cap\mathbb{Q}[Y_1,\cdots,Y_k]$$
In order to define the notion of discriminant variety according to our assumptions, we introduce the notion of \emph{geometric regularity}.
\begin{definition}
  Let $E$ be a subset of the parameters space.\\
  A parametric system $S$ defining a constructible set $\mathcal{C}$ is said to be \emph{geometrically regular} over $E$ iff for all open set $\mathcal{U}\subset E$, $\pi$ restricted to $\pi^{-1}(\mathcal{U})\cap\mathcal{C}$ is an analytic covering.
\end{definition}
The minimal discriminant variety is now defined as follows.
\begin{definition}\cite{rouillier/lazard}
  A \emph{discriminant variety} of the parametric system $S$ is a variety $V$ in the parameters space such that $S$ is \emph{geometrically regular} over $\mathbb{C}^s \setminus V$.
\end{definition}
Among the discriminant varieties we define the \emph{minimal} one:
\begin{definition}\cite{rouillier/lazard}
  The \emph{minimal discriminant variety} of $S$ is the intersection of all the discriminant varieties of $S$.
\end{definition}
For the computation of the minimal discriminant variety, we will assume some properties on the input parametric systems we consider.
\begin{definition}
  \label{gensimp}
   Let $S$ be the parametric system defined by:
$$\left\{ \begin{array}{c}f_1(t,x)=0\\\vdots\\f_n(t,x)=0\end{array}\right.
  \mbox{ and }
  \left\{ \begin{array}{c} g_1(t,x)\neq0\\\vdots\\g_r(t,x)\neq0\end{array}\right.\quad(t,x)\in\mathbb{C}^s\times\mathbb{C}^n$$
    Denoting $\prod_{j=1}^rg_j$ by $g_S$, assume that the ideal in the polynomial ring over the field of fractions of the parameters
    $$I^e=\left< f_1,\cdots,f_n\right>:g_S^\infty \subset\mathbb{Q}(T_1,\cdots,T_s)[X_1,\cdots,X_n]$$
    is radical and 0-dimensional.\\
    Then $S$ is said \emph{generically simple}.
\end{definition}
\begin{remark}
  Note that the ideal $I$ generated by $f_1,\cdots,f_n\subset\mathbb{Q}[T_1,\cdots,T_s,X_1,\cdots,X_n]$ needs neither to be radical nor equidimensional, although it is sufficient to satisfy the hypotheses.
\end{remark}
Moreover, given a parametric system $S$ defined by $f_1=0,\cdots,f_n=0,g_1\neq 0,\cdots,g_r\neq 0$, we introduce these two polynomials:\\
\parbox{0.1\columnwidth}{${}- j_S$\\}\parbox{0.9\columnwidth}{is the determinant of the Jacobian matrix of $f_1,\cdots,f_n$ with respect to the unknowns, of degree denoted by $\delta$}\\
\parbox{0.1\columnwidth}{${}- g_S$}\parbox{0.9\columnwidth}{is the product of the $g_i$ for $1\leq i\leq r$ of degree denoted by $\delta'$}\\
Note that we have $\delta\leq\sum_{i=1}^n d_i-n$ and $\delta'=\sum_{j=1}^rd'_j$.
\subsection*{Main results}
We can now state our main results.
\begin{theorem}
  \label{degree}
  Let $S$ be a \emph{generically simple} parametric system.\\
  Then the total degree of the minimal discriminant variety is bounded by $$d_1\cdots d_n(1+\delta+\delta')$$
\end{theorem}
\begin{theorem}
  \label{reduction}
  Let $S$ be a parametric system \emph{generically simple} defined by $f_1=0,\cdots,f_n=0,g_1\neq 0,\cdots,g_r\neq 0$. Then the union of the varieties defined by the $n+2$ following ideals:\\
- $R$ denotes the ring $\mathbb{Q}[T_1,\cdots,T_s]$\\
  $$\begin{array}{l@{\hspace{2cm}}c@{\hspace{2cm}}c}
    &\left(\left<f_1^h,\cdots,f_n^h,ZX_0g_S^h-1,X_1-1\right>\cap R[X_0]\right)_{|X_0=0}&
    {(\mathcal{I}_1)}\\
    &\vdots&
    {\vdots}\\
    &\left(\left<f_1^h,\cdots,f_n^h,ZX_0g_S^h-1,X_n-1\right>\cap R[X_0]\right)_{|X_0=0}&
    {(\mathcal{I}_n)}\\ \\
    &\left(\left<f_1,\cdots,f_n,g_S-X_{n+1},ZX_{n+1}-1\right>\cap{}R[X_{n+1}]\right)_{|X_{n+1}=0}&
    {(\mathcal{I}_{n+1})}\\ \\
    &\left(\left<f_1,\cdots,f_n,j_S,Zg_S-1\right>\right)\cap R&
    {(\mathcal{I}_{n+2})}\\ \\
  \end{array}$$
  is the minimal discriminant variety of $S$.
\end{theorem}
\begin{corollary}
  \label{complexity}
  A discriminant variety of a \emph{generically simple} parametric system can be computed in:
  $$\sigma^{\mathcal{O}(1)}(d_1\cdots d_n(\delta+\delta'))^{\mathcal{O}(n+s)}$$
  steps on a classical Turing machine. The variable $\sigma$ denotes the maximal binary size of coefficients of $f_1,\cdots,f_n$ and $g_1,\cdots,g_r$.
\end{corollary}
\begin{remark}
  If the system is not \emph{generically simple}, then the the union of the varieties computed is the whole parameter space, which is thus an easy way to check if the initial conditions are verified.
\end{remark}
\begin{remark}
  Any elimination algorithm may actually be used to compute a discriminant variety, which is welcomed when it comes to an effective computation. Among others, Gr\"obner bases with a block ordering \cite{Faugere:2002:NEA,MR2093185}, sparse elimination \cite{Cox-Little-O_Shea/98} or straight-line programs \cite{MR1624458} may lead to efficient computations.
\end{remark}
\begin{remark}
  \label{probatrick}
  If we allow ourself to use the model of a probabilistic bounded Turing Machine, then at the cost of the sparsity of the system, we may replace the computation of $\mathbf{V}(\mathcal{I}_1),\ldots,\mathbf{V}(\mathcal{I}_n)$ by the computation of the variety of:
  $$(\left<f_1^h,\cdots,f_n^h,ZX_0g_S^h-1,\gamma_1X_1+\cdots+\gamma_nX_n-1\right>\cap\mathbb{Q}[T_1,\cdots,T_s][X_0])_{|X_0=0}$$
where $(\gamma_1,\ldots,\gamma_n)$ is chosen randomly in $\{0,\ldots,D-1\}^n$ and $D:=3d_1\cdots d_n$.\footnote{The remark \ref{probatrick} and the corollary \ref{complexity} are proved Section \ref{degree issues}}

\end{remark}
\section{Log-space reduction}
\subsection{Preliminaries}
The goal of this section is to show how to reduce the problem of computing the minimal discriminant variety (the \emph{discriminant problem}) to the \emph{elimination problem}. We know that the \emph{elimination problem} is solvable in polynomial space (\cite{Matera97}). Thus via the reduction we prove that the problem of computing the minimal discriminant variety is solvable in polynomial space.\bigskip

\textsc{Discriminant Function:}\\
\parbox{0.17\columnwidth}{\emph{- Input}:} \parbox{0.82\columnwidth}{$f_1,\cdots,f_n,g_S,j_S\in\mathbb{Q}[T_1,\cdots,T_s,X_1,\cdots,X_n]$}\\
\parbox[h]{0.17\columnwidth}{\emph{- Output}:\\ \\} \parbox[h]{0.82\columnwidth}{$q_{1,1},\cdots,q_{t,u_t}\in\mathbb{Q}[T_1,\cdots,T_s]$ such that  $\cup_{i=1}^t\mathbf{V}(\left<q_{i,1},\cdots,q_{i,u_i}\right>)$ is the minimal discriminant variety.}\\\bigskip

\textsc{Elimination Function: 
}\\
\parbox{0.17\columnwidth}{\emph{- Input}:\\} \parbox{0.82\columnwidth}{$\left\{\begin{array}{l}p_1,\cdots,p_m\in\mathbb{Q}[T_1,\cdots,T_s][X_1,\cdots,X_n];\\T_1,\ldots,T_s\end{array}\right.$}\\
\parbox[h]{0.17\columnwidth}{\emph{- Output}:\\\\} \parbox[h]{0.82\columnwidth}{$q_1,\cdots,q_t\in\mathbb{Q}[T_1,\cdots,T_s]$ such that $\mathbf{V}(\left<q_1,\cdots,q_t\right>)$ is the variety of the elimination ideal $\left<p_1,\cdots,p_m\right>\cap\mathbb{Q}[T_1,\cdots,T_s]$.}\\\bigskip

To achieve the reduction, we will first describe more precisely how the minimal discriminant variety can be decomposed. In \cite{rouillier/lazard}, the authors show that the minimal discriminant variety of a \emph{generically simple} parametric system S is the union of 3 varieties, denoted respectively by $V_{inf}$, $V_{ineq}$ and $V_{crit}$. Let us remind the definitions of these varieties under our assumptions.
\begin{definition}
  \label{def_dv}
  Let $S$ be a \emph{generically simple} parametric system defined by $f_1=0,\cdots,f_{n}=0 \mbox{ and } g_1\neq0,\cdots,g_r\neq0$. The varieties $V_{inf}$,$V_{ineq}$ and $V_{crit}$ of the parameters space are respectively defined as follow:
  \begin{itemize}
    \item[\parbox{0.7cm}{$V_{inf}$}]$=\overline{\pi}(\overline{\mathcal{C}}_S\cap\mathcal{H}_\infty)$\\
      where $\overline{\mathcal{C}}_S$ is the projective closure of the constructible set defined by $S$, and $\mathcal{H}_\infty=(\mathbb{C}^s\times\mathbb{P}_n)\setminus(\mathbb{C}^s\times\mathbb{C}^n)$ is the hypersurface at the infinity.
    \item[\parbox{0.7cm}{$V_{ineq}$}]$=\textbf{V}\left(\left(I_S:g_S^\infty+\left<g_S\right>\right)\cap\mathbb{Q}[T_1,\cdots,T_s]\right)$
    \item[\parbox{0.7cm}{$V_{crit}$}]$=\textbf{V}\left(\left(I_S:g_S^\infty+\left<j_S\right>\right)\cap \mathbb{Q}[T_1,\cdots,T_s]\right)$ 
  \end{itemize}
\end{definition}
\begin{theorem}\cite{rouillier/lazard}
  The \emph{minimal discriminant variety} of a \emph{generically simple} parametric system is the union of $V_{inf}$, $V_{ineq}$ and $V_{crit}$.
\end{theorem}
Geometrically, this theorem characterizes the different varieties in the parameter space over which the \emph{generically simple} parametric system is not \emph{geometrically regular}. More precisely, the theorem means that over the minimal discriminant variety, three types of irregularity may appear. The first one is the intersections of the system of equations with the Jacobian. The second one is the intersection with the inequations. And the last one is the intersection in the projective space of the the hypersurface at the infinity with the projective closure of the parametric system's zeros.

$V_{crit}$ is already directly the solution of an \emph{elimination problem}. This is the component for which the generic radicality condition is needed. We will now focus on reducing the computation of each of the two varieties $V_{inf}$ and $V_{ineq}$ to the \emph{elimination problem}.
\subsection{Reduction of {\large${V_{inf}}$} and correctness}
\label{vinf}
Before going further, it should be clear that the computation of $V_{inf}$ can not be handled by the standard projective elimination methods if we want to certify a singly exponential complexity. All of these methods have no good complexity bounds essentially because of the intersection with the particular hypersurface at the infinity as we will see later. However this doesn't prevent us to use results of the projective elimination theory.

Using the algebraic representation of the projection $\overline\pi$ of \cite{Cox92}, with the notations of the definition \ref{def_dv} we reformulate $V_{inf}$:
\begin{eqnarray*}
  V_{inf}&=&\mathbf{V}\left(\left(\bigcap_{i=1}^n (J_S)_{|X_0=0}:X_i^\infty\right)\cap\mathbb{Q}[T_1,\cdots,T_s]\right)
\end{eqnarray*}
where $J_S:=(I_S:g_S^\infty)^h$. Note that $\overline{\mathcal{C}}_S=\mathbf{V}(J_S)$.

And using the reformulation of the ideal homogenization of \cite{Cox92}, we obtain a formulation of $J_S$ which match explicitly the input of the problem:
\begin{eqnarray*}
  J_S&=&\left<f_1^h,\cdots,f_n^h\right>:{g_S^h}^\infty:X_0^\infty
\end{eqnarray*}
This is however not yet satisfying since this formulation is not trivially reducible to a single elimination problem. The problem here does not come from the saturation by the variables $X_i$ which can be simply handled with the Rabinowitsch trick \cite{MR1512592} of adding the new variable $Z$ and the new equation $ZX_i-1$ to the initial polynomials. Neither is the saturation by $g_S$ a problem since again we may add the equation $Zg_S-1=0$. The complications arise actually from the variable $X_0$. First we have to saturate by $X_0$ and then we have to specialize $X_0$ with $0$ to finally eliminate the variables $X_i$. And it is regrettable since this prevents us to use the usual trick to get rid of the saturation, as we saw in introduction. Moreover we don't want to apply successively two \textsc{Elimination Function} since it could lead us to an exponential space algorithm.\\
Fortunately we manage to sort out this problem by proving that for the variety we want to compute, we can commute the specialization of $X_0$ by $0$ and the elimination, which is remarkable since this operation will allow us to use the Rabinowitsch trick to localize by $X_0$. Note that the commutation step does not alter the computation only because of the particular structure of $V_{inf}$. 
\begin{proposition}
  \label{inf}
  Let $S$ be a parametric system. Then the component $V_{inf}$ of the minimal discriminant variety of $S$ is the union of the varieties defined by the $n$ following ideals for $1\leq i\leq n$:
  $$\left(\left<f_1^h,\cdots,f_n^h,ZX_0g_S^h-1,X_i-1\right>\cap R[X_0]\right)_{|X_0=0}$$
\end{proposition}
\begin{remark}
  Note that the condition \emph{generically simple} is not needed for the reduction of the computation of $V_{inf}$. Moreover the proposition remains true even if the number of equations differs from the number of unknowns.
\end{remark}
The proof of this proposition is based on the three following lemmas. The first one gives some basic useful equalities, where $h_i$ denotes the homogenization by the variable $X_i$ with respect to the variables $X_0,\cdots,\hat{X_i},\cdots,X_n$.
\begin{lemma}\cite{Cox92}
  \label{basics}
  Let $J\subset \mathbb{Q}[T_1,\cdots,T_s][X_0,\cdots,X_n]$ be an ideal homogeneous in $X_0,\cdots,X_n$ and $p$ be a polynomial of $\mathbb{Q}[T_1,\cdots,T_s][X_0,\cdots,X_n]$ also homogeneous in $X_0,\cdots,X_n$. Then for all $0\leq i\leq n$ we have:
  $$(J_{|X_i=1})^{h_i}=J:X_i^\infty$$
  $$(J:p^\infty)_{|X_i=1}=J_{|X_i=1}:p_{|X_i=1}^\infty$$
  $$J:X_i^\infty\cap\mathbb{Q}[T_1,\cdots,T_s]=J_{|X_i=1}\cap\mathbb{Q}[T_1,\cdots,T_s]$$
  and for all $1\leq i\leq n$:
  $$J_{|X_i=1}\cap\mathbb{Q}[T_1,\cdots,T_s][X_0]=(J\cap\mathbb{Q}[T_1,\cdots,T_s][X_0,X_i])_{|X_i=1}$$
\end{lemma}
\begin{proof}
  These are classical results that can be recovered from \cite{Cox92}.
\end{proof}
Now comes the first lemma toward the reduction, which proves essentially that the union of the varieties defined by the elimination ideals of the proposition \ref{inf} contains $V_{inf}$.
\begin{lemma}
  \label{inf1}
   Let $J$ be an ideal of $\mathbb{Q}[T_1,\cdots,T_s][X_0,\cdots,X_n]$ homogeneous in $X_0,\cdots,X_n$. Then, for all $1\leq i\leq n$ we have:
\begin{eqnarray*}
  &(J\cap\mathbb{Q}[T_1,\cdots,T_s][X_0,X_i])_{|X_0=0,X_i=1}&\\
  &\cap&\\
  &(J_{|X_0=0}:X_i^\infty)\cap\mathbb{Q}[T_1,\cdots,T_s]&
\end{eqnarray*}
\end{lemma}
\begin{proof}
  Let $p\in(J\cap\mathbb{Q}[T_1,\cdots,T_s][X_0,X_i])_{|X_0=0,X_i=1}$. The polynomial $p$ is homogeneous in $X_0,\ldots,X_n$ since it depends only on the variables $T_1,\ldots,T_s$. Thus with the notations of the lemma \ref{basics}, we have $p\in ((J_{|X_0=0})_{|X_i=1})^{h_i}$. And $J_{|X_0=0}$ being homogeneous in $X_0,\cdots,X_n$, one can apply the first equality of Lemma \ref{basics} to deduce $p\in J_{|X_0=0}:X_i^\infty$ which proves the desired result.
\end{proof}
And finally comes the keystone lemma related to the proposition, proving the reciprocal inclusion.
\begin{lemma}
  \label{inf2}
   Let $J$ be an ideal of $\mathbb{Q}[T_1,\cdots,T_s][X_0,\cdots,X_n]$ homogeneous in $X_0,\cdots,X_n$. Then, for all $1\leq i\leq n$, we have:
\begin{eqnarray*}
  &\sqrt{(J\cap\mathbb{Q}[T_1,\cdots,T_s][X_0,X_i])_{|X_0=0,X_i=1}}&\\
  &\cup&\\
  &\bigcap_{j=1}^n(J_{|X_0=0}:X_j^\infty)\cap\mathbb{Q}[T_1,\cdots,T_s]&
\end{eqnarray*}
\end{lemma}
\begin{proof}
  Let $p\in\bigcap_{j=1}^n(J_{|X_0=0}:X_j^\infty)\cap\mathbb{Q}[T_1,\cdots,T_s]$. By definition there exist $q_1,\cdots,q_n \in \mathbb{Q}[T_1,\cdots,T_s][X_0,\cdots,X_n]$ and $k_1,\cdots,k_n\in\mathbb{N}$ such that:
$$\left\{\begin{array}{rcl}
  p_1&:=&pX_1^{k_1}+X_0q_1\\
  &\vdots&\\
  p_n&:=&pX_n^{k_n}+X_0q_n\\
\end{array}\right.\in J$$
Since the part of $p_i$ of degree $k_i$ in $X_0,\cdots,X_n$ belongs also to $J$, we can assume that $p_1,\cdots,p_n$ are homogeneous in $X_0,\cdots,X_n$. Thus, we have in particular:
$$\deg_{X_1,\cdots,X_n}(q_j)<k_j$$
Now we fix a total degree term order $<_{X}$ on the variables $X_1,\cdots,X_n$. Let $K$ denote the field $\mathbb{Q}(T_1,\cdots,T_s,X_0)$ and consider $p_1,\cdots,p_n$ as polynomials of $K[X_1,\cdots,X_n]$. Denoting by $\mathcal{J}$ the ideal they generate, it follows immediately that
$$G:=\{p_1,\cdots,p_n\}$$
form a Gr\"obner basis of $\mathcal{J}$ with respect to $<_{X}$ since the $p_i$ have disjoint head terms.
Let $i$ be an integer between $1$ and $n$. We first show how to prove the lemma when we have a polynomial of $\mathcal{J}$ such that:\\
\begin{tabular*}{\columnwidth}{l@{\extracolsep\fill}r}
  - it is univariate in $X_i$&(1)\\
  - it has all its coefficients in $\mathbb{Q}[T_1,\cdots,T_s,X_0]$&(2)\\
  - its head coefficient is a power of $p$&(3)\\
\end{tabular*}
Assume for a while that $r_{i}$ is such a polynomial, $d_{X_i}$ being its degree in $X_i$. It follows indeed that $r_{i}\in\mathcal{J}^c$ the contraction ideal of $\mathcal{J}$. And since $p=lcm\{HC(g) | g\in G\}$ we have \cite{MR1213453}:
$$\mathcal{J}^c=\left<G\right>:p^\infty$$
meaning that for some $k\in\mathbb{N}$, $p^kr_{i}\in\left<G\right>\subset J$. Finally $J$ is homogeneous so that $\tilde{r_i}$, the part of degree $d_{X_i}$ of $p^kr_{i}$, belongs to $J\cap\mathbb{Q}[T_1,\cdots,T_s][X_0,X_i]$ and can be written as:
$$ \tilde{r_i}=p^lX_i^{d_i}+X_0q$$
with $l\in\mathbb{N}$ and $q\in\mathbb{Q}[T_1,\cdots,T_s][X_0,X_i]$, which is an equivalent way of writing
$$p\in\sqrt{(J\cap\mathbb{Q}[T_1,\cdots,T_s][X_0,X_i])_{|X_0=0,X_i=1}}$$
It remains us to show the existence of a polynomial satisfying (1),(2) and (3). To carry out this problem, we first notice that $\mathcal{J}$ is zero-dimensional in $K[X_1,\cdots,X_n]$ since the set of the head terms of its Gr\"obner basis contains a pure power of each variable $X_i$. So, we may consider the finite dimensional $K-$space vector $\mathcal{A}=K[X_1,\cdots,X_n]/\mathcal{J}$ along with $\mathbf{e}$ the monomial basis of $\mathcal{A}$ induced by $G$. More precisely, denoting by $\overline{x}$ the class of $x$ in $\mathcal{A}$, we define see $\mathbf{e}$ as the set of $\overline{e_j}$ for $1\leq j\leq D:=\dim(\mathcal{A})$ such that $e_j$ is a term of $K[X_1,\cdots,X_n]$ not multiple of any head term of $G$. Finally we denote by $S$ the multiplicatively closed set $\{p^k,k\in\mathbb{N}\}$. We will follow a classical method to exhibit a monic univariate polynomial from a zero-dimensional ideal, with coefficients in $K$. And with results of \cite{MR1213453} we ensure that its coefficients are not only in $K$ but rather in the ring $S^{-1}\mathbb{Q}[T_1,\cdots,T_s,X_0]\subset K$.\\
Let us introduce the classical linear application of multiplication by $X_i$:
 $$\begin{array}{cccc}
   \Phi_i:&\mathcal{A}&\rightarrow&\mathcal{A}\\
   &\overline{q}&\mapsto&\overline{X_iq}
 \end{array}$$
 Then we note $M_i$ the matrix of $\Phi_i$ in the base $\mathbf{e}$:
 $$M_i=\begin{array}{cccc}
   &\overline{X_ie_1}&\cdots&\overline{X_ie_D}\\
   \begin{array}{c}
     \overline{e_1}\\
     \vdots\\
     \overline{e_D}
   \end{array}
   &
   \multicolumn{3}{c}{
   \left(
   \begin{array}{ccc}
     &&\\
     \quad&c_{k,l}&\quad\\
     &&\\
   \end{array}\right)}
 \end{array}
  $$
  we notice that the coefficients of $M_i$ come from the reduction of the monomials $X_ie_l$ for $1\leq l\leq D$ by the Gr\"obner basis $G$. And as we can see in \cite{MR1213453}, this kind of reduction only involves division by the head coefficients of $G$, such that:
  $$\overline{X_ie_l}=c_{1,l}\overline{e_{1,l}}+\cdots+c_{D,l}\overline{e_{D,l}}$$
  with $c_{1,l},\cdots,c_{D,l}$ not only in $K$ but more precisely in the ring $S^{-1}\mathbb{Q}[T_1,\cdots,T_s,X_0]\subset K$ where $S=\{p^k,k\in\mathbb{N}\}$. As a straightforward consequence, if we denote by $\mathcal{P}_i$ the monic characteristic polynomial of $M_i$ in the new variable $U$, we have $\mathcal{P}_i\in S^{-1}\mathbb{Q}[T_1,\cdots,T_s,X_0][U]$. Besides by the Cayley-Hamilton's theorem, $\mathcal{P}_i$ applied to the variable $X_i$ is the null element of $\mathcal{A}$, meaning that $\mathcal{P}_i(X_i)$ belongs to $\mathcal{J}$ and may be written as:
  $$\mathcal{P}_i(X_i)=X_i^{D}+C_{D-1}X_i^{D-1}+\cdots+C_0$$
  with $C_k\in S^{-1}\mathbb{Q}[T_1,\cdots,T_s,X_0]$ for $1\leq k\leq D-1$. Finally, for some $k'\in\mathbb{N}$ we have $$r_i:=p^{k'}\mathcal{P}_i(X_i)\in\mathcal{J}\cap\mathbb{Q}[T_1,\cdots,T_s,X_0][X_i]$$ which satisfies all the conditions we wanted to achieve the demonstration.
\end{proof}
Finally, a proper combination of the lemmas proves the proposition \ref{inf}.
\subsection{Reduction of {\large${V_{ineq}}$} and correctness}
As to bound the computation of the variety induced by the inequations $$V_{ineq}=\textbf{V}\left(\left(I_S:g_S^\infty+\left<g_S\right>\right)\cap\mathbb{Q}[T_1,\cdots,T_s]\right)$$ the direct approach consists first in performing a saturation and then in using the output along with $g_S$ as the input of an elimination algorithm. However this method may not have a single exponential bound on the time complexity in the worst case. Hence both of these algorithms may use a polynomial space in the size of the input, which could finally cost an exponential space if no more care is taken.

In this section we show how to bypass the problem, notably by relaxing the condition on the output and allowing some components of $V_{inf}$ to mix in.
\begin{proposition}
  \label{ineq}
  Let $S$ be a parametric system. If we denote by $W_{ineq}\subset\mathbb{C}^s$ the variety defined by the following ideal:
  \begin{eqnarray*}
    \lefteqn{(\left<f_1,\cdots,f_n,g_S-X_{n+1},ZX_{n+1}-1\right>}&&\\
    &\hspace{3cm}\cap{}\mathbb{Q}[T_1,\cdots,T_s][X_{n+1}])_{|X_{n+1}=0}&
\end{eqnarray*}
  then the following inclusions chain holds:
  $$V_{ineq}\subset W_{ineq}\subset V_{ineq}\cup V_{inf}$$
\end{proposition}
The first step to prove this proposition is to delay the saturation.
\begin{lemma}
  Let $p_1,\cdots,p_m,q,r\in\mathbb{Q}[Y_1,\cdots,Y_k]$. Let us fix $<$ a term order and assume that the head monomial of $q$ shares no variable in common with the monomials of $p_1,\cdots,p_m,r$.   Then we have the following equality:
  $$\left<p_1,\cdots,p_m\right>:r^\infty+\left<q\right>=\left<p_1,\cdots,p_m,q\right>:r^\infty$$
\end{lemma}
\begin{proof}
  The inclusion from left to right is trivial. For the other inclusion, let $p\in \left<p_1,\cdots,p_m,q\right>:r^\infty$. Denoting by $M$ the head monomial of $q$ with respect to $<$, we obtain by division:
  \begin{eqnarray}
    p=p'+qt &&  p',t\in\mathbb{Q}[Y_1,\cdots,Y_k]
  \end{eqnarray} 
  such that no monomial of $p'$ is multiple of $M$. It remains to show that $p'$ belongs to $\left<p_1,\cdots,p_m\right>:r^\infty$ and the proof will be complete. By hypothesis, we know that there exists $l>0$ and $c_1,\cdots,c_m,c\in \mathbb{Q}[Y_1,\cdots,Y_k]$ such that:
  $$r^lp'=c_1p_1+\cdots+c_mp_m+cq$$
We divide each of the $c_i$ by $q$ as in (1) and denote by $c_i'$ the remainder of the division. We thus obtain:
$$r^lp'-c_1'p_1-\cdots-c_m'p_m=c'q$$
with $c'\in\mathbb{Q}[Y_1,\cdots,Y_k]$. We remark that the polynomial on the left part of the equality has no monomial which is multiple of $M$, while the head monomial of the right part of the equality is $M$ times the head monomial of $c'$, which means $c'=0$ and this achieves the proof.
\end{proof}
\begin{corollary}
  Let $f_1,\cdots,f_n,g$ be some polynomials of $\mathbb{Q}[T_1,\cdots,T_s][X_1,\cdots,X_n]$. Then:
  $$\left<f_1,\cdots,f_n\right>:g^\infty+\left<g-X_{n+1}\right>=\left<f_1,\cdots,f_n,g-X_{n+1}\right>:X_{n+1}^\infty$$
\end{corollary}
Thanks to this result, we can now reformulate $V_{ineq}$ as being the variety of:
$$\left(\left<f_1,\cdots,f_n,g_S-X_{n+1}\right>:X_{n+1}^\infty\right)_{|X_{n+1}=0}\cap\mathbb{Q}[T_1,\cdots,T_s]$$
The reduction is not yet complete and we encounter here the same problem we had for the computation of $V_{inf}$, that is the saturation by $X_{n+1}$ before the specialization of $X_{n+1}$ by $0$. This is just fine since the lemmas \ref{basics},\ref{inf1} and \ref{inf2} provide us tools to handle it, even if they do not completely solve the problem yet. 

For the first inclusion, we note:
$$I^S:=\left<f_1,\cdots,f_n,g_S-X_{n+1}\right>:X_{n+1}^{\infty}$$
it follows that the varieties of the proposition \ref{ineq} rewrite as:
\begin{eqnarray*}
  V_{ineq}&=&\mathbf{V}\left(I^S_{|X_{n+1}=0}\cap\mathbb{Q}[T_1,\cdots,T_s]\right)\\
  W_{ineq}&=&\mathbf{V}\left((I^S\cap\mathbb{Q}[T_1,\cdots,T_s][X_{n+1}])_{|X_{n+1}=0}\right)
\end{eqnarray*}
and we show easily:
$$(I^S\cap\mathbb{Q}[T_1,\cdots,T_s][X_{n+1}])_{|X_{n+1}=0}\subset I^S_{|X_{n+1}=0}\cap\mathbb{Q}[T_1,\cdots,T_s]
$$
which, in term of varieties, proves the first inclusion of the proposition \ref{ineq}.

For the other inclusion we will mainly use the lemma \ref{inf2}. For this, we introduce the homogenization variable $X_{0}$, and denote with the exponent $h$ the homogenization by $X_{0}$ with respect to $X_0,\cdots,X_{n+1}$. We need also the following classical lemma, which dissociates the affine part from the component at the infinity of a homogeneous ideal.
\begin{lemma}\cite{Cox92}
  \label{split}
  Let $J\in\mathbb{Q}[T_1,\cdots,T_S][X_{0},\cdots,X_{n+1}]$ an ideal  homogeneous in $X_0,\cdots,X_{n+1}$. Then the following equality holds:
  $$\sqrt{J}=\sqrt{J+\left<X_0\right>}\cap \sqrt{J:X_0^\infty}$$
\end{lemma}
  In term of varieties, the equality follows from the observation that $\mathbf{V}(J)$ is the union of $\mathbf{V}(J)\cap\mathcal{H}_\infty$ and $\overline{\mathbf{V}(J)\setminus\mathcal{H}_\infty}$.\\
We now homogenize $I^S$ by $X_0$ and we get:
$$W_{ineq}=\mathbf{V}({I^S}^h\cap\mathbb{Q}[T_1,\cdots,T_s][X_0,X_{n+1}]_{|X_{n+1}=0,X_0=1})$$
Using lemma \ref{inf2}, we get directly:
$$W_{ineq}\subset\mathbf{V}\left(\bigcap_{j=0}^n({(I^S}^h)_{|X_{n+1}=0}:X_j^\infty)\cap\mathbb{Q}[T_1,\cdots,T_s]\right)$$
Then we show that $({I^S}^h)_{|X_{n+1}=0}$ contains an ideal which begins to look like what we want:
\begin{eqnarray*}
{I^S}^h&=&\left(\left<f_1,\cdots,f_n\right>:g_S^\infty+\left<g_S-X_{n+1}\right>\right)^h\\
({I^S}^h)_{|X_{n+1}=0}&\supset&\left<f_1^h,\cdots,f_n^h\right>:{g_S^h}^\infty:X_0^\infty+\left<X_0g_S^h\right>\\
&\supset&J_S+\left<X_0g_S^h\right>
\end{eqnarray*}
Then, the lemma \ref{split} allows us to split the ideal $J_S+\left<X_0g_S^h\right>$ in:
$$
\begin{array}{c@{}c@{}c@{}c}
  \sqrt{J_S+\left<X_0g_S^h\right>}={}&\underbrace{\sqrt{J_S+\left<X_0g_S^h\right>+\left<X_0\right>}}&{}\cap{}&\underbrace{\sqrt{(J_S+\left<X_0g_S^h\right>):X_0^\infty}}\\
  &I_1&&I_2
\end{array}  
  $$
  such that we now have the following inclusion:
  $$W_{ineq}\subset\mathbf{V}\left(\bigcap_{j=0}^n(I_1:X_j^\infty)\cap (I_2:X_j^\infty)\cap\mathbb{Q}[T_1,\cdots,T_s]\right)$$
  From there, denoting again $\mathbb{Q}[T_1,\ldots,T_s]$ by $R$, we remark for $0\leq j\leq n$:
\begin{eqnarray*}
  I_1:X_j^\infty\cap R&\supset& (J_S)_{|X_0=0}:X_j^\infty\cap R
\end{eqnarray*}
And:
\begin{eqnarray*}
  I_2:X_j^\infty\cap R&\supset&\left(J_S+\left<g_S^h\right>\right):X_j^\infty:X_0^\infty \cap R\\
  &\supset&\left((J_S)_{|X_0=1}+\left<g_S\right>\right):X_j^\infty \cap R\\
  &\supset&I^S:X_j^\infty\cap R\\
  &\supset&I^S\cap R
\end{eqnarray*}
Which allows us to conclude with:\\
$$\begin{array}{l@{}c@{}l}
  W_{ineq}&{}\subset{}&\mathbf{V}\left(I^S\cap\bigcap_{j=0}^n\left({J_S}_{|X_0=0}):X_j^\infty\right)\cap\mathbb{Q}[T_1,\cdots,T_s]\right)\\\\
  &{}\subset{}&V_{ineq}\cup V_{inf}
\end{array}$$
This proves the theorem \ref{reduction}.
\section{Degree issues}
\label{degree issues}
The study of the degree of the minimal discriminant variety relies strongly on the Bezout-Inequality \cite{TCS::Heintz1983,MR1644323}. What we call degree of an ideal $I$ (resp. a variety $V$) and denote $\deg(I)$ (resp. $\deg(V)$) is the sum of the degrees of the prime ideals associated to $\sqrt{I}$ (resp. the sum of the degrees of the irreducible components of $V$). With this definition, from \cite{TCS::Heintz1983,MR1644323} we have for $I,J\subset\mathbb{Q}[T_1,\ldots,T_s,X_0,\ldots,X_n]$ and $f\in\mathbb{Q}[T_1,\ldots,T_s,X_0,\ldots,X_n]$:
\begin{eqnarray*}
  \deg(I+J)&\leq& \deg(I)\deg(J)\\
  \deg(I:f^\infty)&\leq& \deg(I)\\
  \deg(I\cap\mathbb{Q}[T_1,\ldots,T_s])&\leq&\deg(I)\\
  \deg(I)&=&\deg(\mathbf{V}(I))
\end{eqnarray*}
\subsection*{Degree of $V_{inf}$}
Here we use the prime decomposition of $\sqrt{J_S}$ to bound the degree of $V_{inf}$. This decomposition will also allow us to prove Remark \ref{probatrick}.

First we remind that from proposition \ref{inf}:
$$V_{inf}=\mathbf{V}\left(\left(\left(\bigcap_{i=1}^n J_S:X_i^\infty\right)\cap\mathbb{Q}[T_1,\cdots,T_s,X_0]\right)_{|X_0=0}\right)$$
where $J_S=\left<f_1^h,\cdots,f_n^h\right>:{g_S^h}^\infty:X_0^\infty$

Continuing with the properties of the degree we have: 
$$\deg(J_S)\leq \deg\left(\left<f_1^h,\cdots,f_n^h\right>\right)\leq d_1\cdots d_n$$
Let $\mathfrak{P}_1,\ldots,\mathfrak{P}_k$ be the prime ideals associated to $\sqrt{J_S}$. Then we have:
$$ \mathfrak{P}_1\cap\cdots\cap\mathfrak{P}_k=\sqrt{J_S} $$
$$\deg(\mathfrak{P}_1)+\cdots+\deg(\mathfrak{P}_k)\leq d_1\cdots d_n$$
Now let denote by $\lambda_1,\ldots,\lambda_j$ the indices of the prime ideal which do not contain any power of $X_i$ for some $1\leq i\leq n$. It follows that:
$$ \bigcap_{i=1}^n\sqrt{J_s:X_i^\infty}=\mathfrak{P}_{\lambda_1}\cap\cdots\cap\mathfrak{P}_{\lambda_j}$$
such that 
$$
\begin{array}{l@{}c@{}l}
  \deg(V_{inf})&{}={}&\deg\left(\bigcap_{i=1}^n\sqrt{J_s:X_i^\infty}\cap\mathbb{Q}[T_1,\cdots,T_s,X_0]_{|X_0=0}\right)\\\\
  &{}\leq{}& d_1\cdots d_n
\end{array}
$$
We use the decomposition of $\sqrt{J_S}$ to prove the remark \ref{probatrick}.\bigskip\\
\begin{proof} (of Remark \ref{probatrick})\\
   We extend Lemma \ref{basics}, where we replace $X_i$ by a homogeneous linear form in $X_0,\cdots,X_n$, which leads to the following property. If $J$ is an ideal of $\mathbb{Q}[T_1,\cdots,T_s][X_0,\cdots,X_n]$ homogeneous in $X_0,\cdots,X_n$, and $L\in\mathbb{Q}[X_0,\cdots,X_n]$ is a homogeneous linear form in $X_0,\cdots,X_n$, then:
$$ J:L^\infty\cap\mathbb{Q}[T_1,\cdots,T_s,X_0]=\left(J+\left<L-1\right>\right)\cap\mathbb{Q}[T_1,\cdots,T_s,X_0]$$
  From there, we know that the prime ideals which contain a power of $X_i$ for all $1\leq i\leq n$ contain in fact all the homogeneous linear forms of $\mathbb{Q}[X_0,\cdots,X_n]$. Let denote by $E$ the $\mathbb{Q}$-spacevector of homogeneous linear forms of $\mathbb{Q}[X_0,\cdots,X_n]$. Thus we have for all $L\in E$:
  $$\sqrt{{J_S}:L^\infty}=\bigcap_{i=1}^j\mathfrak{P}_{\lambda_i}:L^\infty$$
  Let $B$ denote the bounded lattice $\{0,\ldots,D-1\}^n$ of $E$, where $D=3d_1\cdots d_n$. And $A$ be defined by:
  $$A:=\bigcup_{i=1}^j(\mathfrak{P}_{\lambda_i}\cap E)$$
  Such that for $L\in B\setminus A$, we have:\\
  \parbox{\columnwidth}{\centering$\bigcap_{i=1}^k\mathfrak{P}_i:L^\infty=\bigcap_{i=1}^j\mathfrak{P}_{\lambda_i}$}\bigskip\\
  And since each $\mathfrak{P}_{\lambda_i}\cap E$ is a strict linear subspace of $E$, it follows that $A$ is the union of $j\leq \prod_{i=1}^nd_i=\frac{D}{3}$ strict linear subspaces of $E$
  . Each $\mathfrak{P}_{\lambda_i}\cap E$ intersects the lattice $B$ in at most $D^{n-1}$ points. Thus the probability of choosing $L$ in $B\cap A$ is $\frac{|B\cap A|}{|B|}\leq \frac{1}{3}$. And for all $L\in B\setminus A$ we have:
  \begin{eqnarray*}
    V_{inf}&=&\mathbf{V}\left(\bigcap_{i=1}^k\mathfrak{P}_i:L^\infty\cap\mathbb{Q}[T_1,\cdots,T_s,X_0]_{|X_0=0}\right)\\
    &=&\mathbf{V}\left( \left(\bigcap_{i=1}^k\mathfrak{P}_i+\left<L-1\right>\right)\mathbb{Q}[T_1,\cdots,T_s,X_0]_{|X_0=0}\right)\\
    &=&\lefteqn{\mathbf{V}((\left<f_1^h,\cdots,f_n^h,ZX_0g_S^h-1,L-1\right>}\\
   &&\hspace{3cm}\cap\mathbb{Q}[T_1,\cdots,T_s,X_0])_{|X_0=0})
 \end{eqnarray*}
\end{proof}
\subsection*{Degree of $V_{ineq}$ and $V_{crit}$}
The degree of the two other components are obtained easily. By definition:
\begin{eqnarray*}
  V_{ineq}&=&\mathbf{V}\left((\left<f_1,\cdots,f_n\right>:g_S^\infty+\left<g_S\right>)\cap\mathbb{Q}[T_1,\cdots,T_s]\right)\\
  V_{crit}&=&\mathbf{V}\left((\left<f_1,\cdots,f_n\right>:g_S^\infty+\left<j_S\right>)\mathbb{Q}[T_1,\cdots,T_s]\right)
\end{eqnarray*}
Thus with the properties of the degree, we have respectively:
\begin{eqnarray*}
  \deg(V_{ineq})&\leq&d_1\cdots d_n\delta'\\
  \deg(V_{crit})&\leq&d_1\cdots d_n\delta\\
\end{eqnarray*}
Hence we proved the theorem \ref{degree}.

\subsection*{Degree of representation of the elimination}
To compute the \textsc{Elimination Function} in a deterministic way, we follow the ideas of \cite{oai:CiteSeerPSU:552837} which uses the affine effective Nullstellensatz to reduce the problem to a linear algebra system of non homogeneous linear form. One could use the ideas of \cite{MR1624458,Giusti-Heintz/93,GiustiSchost:1999} to perform this elimination, whose complexity bounds rely on the projective effective Nullstellensatz of \cite{Lazard/77}. However these bounds only hold in a bounded probabilistic Turing machine.

Here we will use the Brownawell's prime power version of Nullstellensatz (see \cite{MR1653279}), which is a variant of the affine effective Nullstellensatz:
\begin{theorem}\cite{MR1653279}
  Let $J\subset k[x_0,\cdots,x_n]$ be an ideal generated by $m$ homogeneous polynomial of respective degrees $d_2\geq\cdots\geq d_m\geq d_1$ and $\mathfrak{M}=\left<x_0,\cdots,x_n\right>$. Then there are prime ideal $\mathfrak{P}_1,\cdots,\mathfrak{P}_r$ containing $J$ and positive integers $e_0,\cdots,e_r$ such that:
  $$\mathfrak{M}^{e_0}\mathfrak{P}_1^{e_1}\cdots\mathfrak{P_r}^{e_r}\subset J\mbox{, and}$$
  $$e_0+\sum_{i=1}^re_i\deg(\mathfrak{P}_i)\leq (3/2)^\mu d_1\cdots d_\mu$$
  where $\mu=\min(m,n)$
\end{theorem}
Using the proposition 3 of \cite{TCS::Heintz1983}, we know that if $\mathfrak{P}$ is a prime ideal, then there is $n+1$ polynomials $f_1,\cdots,f_{n+1}$ such that:
$$ \mathbf{V}(f_1,\cdots,f_{n+1})=\mathbf{V}(\mathfrak{P})$$
with $\deg(f_i)\leq\deg(\mathfrak{P})$ for all $1\leq i\leq n+1$

Thus we deduce the following:
\begin{proposition}
  Let $I\subset\mathbb{Q}[T_1,\cdots,T_s][X_1,\cdots,X_n]$ generated by $f_1,\cdots,f_m$ indexed such that their degrees satisfy $d_2\geq\cdots\geq d_m\geq d_1$. Then, with $\mu=\min(m,n)$ we introduce:
  $$ F:=\left\{\sum_{i=1}^mg_if_i |
  \begin{array}{c}g_i\in\mathbb{Q}[T_1,\cdots,T_s][X_1,\cdots,X_n]\\
    and\\
    \deg(g_if_i)\leq (3/2)^\mu d_1\cdots d_\mu
  \end{array}
    \right\}$$
  Then we have:
  $$ \mathbf{V}(I\cap\mathbb{Q}[T_1,\cdots,T_s])=\mathbf{V}( F\cap\mathbb{Q}[T_1,\cdots,T_s])$$
\end{proposition}
\begin{proof}
  We homogenize the polynomials $f_1,\cdots,f_m$ by $H$ with respect to $T_1,\cdots,T_s,X_1,\cdots,X_n$, and denote by $J$ the ideal they generate. Then with $\mathfrak{P}_1,\cdots,\mathfrak{P}_r$ being prime ideals containing $J$ and verifying the theorem of Brownawell, it follows that the result holds when intersecting $J$ and $\mathfrak{P}_1,\cdots,\mathfrak{P}_r$  by $\mathbb{Q}[T_1,\cdots,T_s,H]$. Finally we use the Heintz's proposition reminded above on each $\mathfrak{P}_i$ and specialize $H$ by $1$ to conclude.  
\end{proof}
Now consider the coefficients of the polynomials $g_1,..,g_m,g$ as unknowns. Assume furthermore that $g_1,\cdots,g_m$ contains all the monomials in $T_1,\cdots,T_s,X_1,\cdots,X_n$ of degree less or equal to $(3/2)^\mu d_1\cdots d_\mu$, and that $g$ contains the monomials in $T_1,\cdots,T_s$ only. Thus, finding the coefficients satisfying the formula:
$$\sum_{i=1}^mg_if_i-g=0$$
reduces to the problem of finding null space generators of a matrix of size lower or equal to $$(m+1)((3/2)^\mu d_1\cdots d_\mu)^{\mathcal(n+s)}\times((3/2)^\mu d_1\cdots d_\mu)^{\mathcal(n+s)}$$ Hence the complexity of the corollary \ref{complexity} follows.

\section{Example}
We show here an example of minimal discriminant varieties application in our framework. It will allow us to prove that the real parametrization of the Enneper surface matches its real implicit form. In \cite{conf/adg/Dolzmann98} the author solves this problem with a combination \verb?REDLOG?,\verb?QEPCAD? and \verb?QERRC?. Through the process, he has to simplify formulas whose textual representation contains approximatively 500 000 characters. We will see that our framework allows us to use \emph{minimal} discriminant varieties to solve this problem. Notably, this allows us to keep formulas small. The following computations are done with the \verb?Maple? package \verb?DV?, which uses \verb?FGb? to carry out the elimination function. We also use the factorization functions of \verb?Maple? to take the square-free part of the polynomials given in the input, and to simplify the output. Finally \verb?RS? and the \verb?Maple? package \verb?RAG? allows us to treat the discriminant varieties we compute. All these software are available in the Salsa Software Suite \cite{SALSA}.\\

When $E$ and $F$ are two lists of polynomials, $T$ a list of parameters and $X$ a list of unknowns, we denote by
$$\mathbf{DV}(E,F,T,X)$$
the discriminant variety of the parametric system $S: (p=0)_{ p\in E}\wedge (q\neq 0)_{q\in F}$.

\subsection{Definition of the Enneper surface}
The real Enneper surface $\mathcal{E}\subset\mathbb{R}^3$ has a parametric definition:\\
$$\mathcal{E}=\left\{(x(u,v),y(u,v),z(u,v)) \quad|\quad (u,v)\in \mathbb{R}^2\right\}$$
$$\begin{array}{rcl}
  x(u,v) &=& 3u+3uv^2-u^3\\
  y(u,v) &=& 3v+3u^2v-v^3\\
  z(u,v) &=& 3u^2-3v^2
\end{array}$$

We will also consider the graph of the Enneper surface $\mathcal{E}_{g}\subset\mathbb{R}^5$ defined as follows: 
$$\mathcal{E}_{g}=\left\{(x(u,v),y(u,v),z(u,v),u,v) \quad|\quad (u,v)\in \mathbb{R}^2\right\}$$

Beside, a Gröbner basis computation returns easily its implicit Zarisky closure $\overline{\mathcal{E}}$ \cite{Cox92,conf/adg/Dolzmann98}:
$$\overline{\mathcal{E}}=\left\{(x,y,z)\in\mathbb{R}^3 \quad|\quad p(x,y,z)=0\right\}$$
$$\begin{array}{rcl}
p(x,y,z)&=&-19683x^6+59049x^4y^2-10935x^4z^3-118098x^4z^2+59049x^4z-59049x^2y^4\\
&&-56862x^2y^2z^3-118098x^2y^2z-1296x^2z^6-34992x^2z^5-174960x^2z^4\\
&&+314928x^2z^3+19683y^6-10935y^4z^3+118098y^4z^2+59049y^4z+1296y^2z^6\\
&&-34992y^2z^5+174960y^2z^4+314928y^2z^3+64z^9-10368z^7+419904z^5
\end{array}$$

\subsection{Discriminant varieties}
The main idea to compare $\mathcal{E_g}$ and $\overline{\mathcal{E}}$ is in a first step to compute the union of their discriminant varieties, $V$. In a second step we compare $\mathcal{E}_g$ and $\overline{\mathcal{E}}$ on a finite number of well chosen test points outside of $V$. Finally, the properties of the discriminant variety ensure us that the result of our comparison on these test points holds for every points outside of $V$.\\

More precisely, $\mathcal{E}_g$ and $\overline{\mathcal{E}}$ are both algebraic varieties of dimension 2. Thus we choose a common subset of 2 variables, $x$ and $y$ for example, which will be the \emph{parameters} for the two discriminant varieties:
$$\begin{array}{l@{\ }c@{\ ,\ }c@{{\ },{\ }}c@{{\ },{\ }}c@{{\ })}}
  V_1^{xy}:=\mathbf{DV}(&[x-x(u,v),y-y(u,v),z-z(u,v)]&[\ ]&[x,y]&[z,u,v]\\
  V_2^{xy}:=\mathbf{DV}(&[p(x,y,z)]&[\ ]&[x,y]&[z]
\end{array}
$$
The number of equations equals the number of unknowns in both case and our algorithm returns a non trivial variety for both systems. This ensures us that the two systems are \emph{generically simple}.
Here are the results of the computations, which lasted less than 1 second on a 2.8 GHz Intel Pentium cpu:\\
$$\begin{array}{ll}
  V_1^{xy}=&\mathbf{V}(y^6+60y^4+768y^2-4096+3x^2y^4-312x^2y^2+768x^2+3x^4y^2+60x^4+x^6)\\
  &\cup\mathbf{V}(x^6+48x^4+3x^4y^2-336x^2y^2+3x^2y^4+768x^2+4096+768y^2+48y^4+y^6)\\\\
  V_2^{xy}=&\mathbf{V}(y^6+60y^4+768y^2-4096+3x^2y^4-312x^2y^2+768x^2+3x^4y^2+60x^4+x^6)\\
  &\cup\mathbf{V}(x^6+48x^4+3x^4y^2-336x^2y^2+3x^2y^4+768x^2+4096+768y^2+48y^4+y^6)\\
  &\cup\mathbf{V}(x-y)\cup \mathbf{V}(y)\cup \mathbf{V}(x+y)\cup \mathbf{V}(x)\\
\end{array}
$$

We denote by $\pi_{xy}:\mathbb{R}^3\rightarrow \mathbb{R}^2$ the canonical projection. Then the properties of the discriminant variety ensure us that for each connected component $\mathcal{C}$ of $\mathbb{R}^2\setminus (V_1^{xy}\cup V_2^{xy})$, $(\pi_{xy}^{-1}(\mathcal{C})\cap\mathcal{E},\pi_{xy})$ and $(\pi_{xy}^{-1}(\mathcal{C})\cap\overline{\mathcal{E}},\pi_{xy})$ are both analytic covering. Moreover, $\mathcal{E}\subset\overline{\mathcal{E}}$. Thus if $\mathcal{C}$ is a connected component of $\mathbb{R}^2\setminus (V_1^{xy}\cup V_2^{xy})$, we get the following property:
$$
\exists p\in \mathcal{C}, \pi_{xy}^{-1}(p)\cap\mathcal{E}=\pi_{xy}^{-1}(p)\cap\overline{\mathcal{E}} \Longleftrightarrow \forall p\in \mathcal{C}, \pi_{xy}^{-1}(p)\cap\mathcal{E}=\pi_{xy}^{-1}(p)\cap\overline{\mathcal{E}}
$$

This allows us to prove that $\mathcal{E}$ and $\overline{\mathcal{E}}$ are equal above $\mathbb{R}^2\setminus (V_1^{xy}\cup V_2^{xy})$: we take one point $p$ in each connected component of $\mathbb{R}^2\setminus (V_1^{xy}\cup V_2^{xy})$, and check that the number of real solutions of $\pi_{xy}^{-1}(p)\cap\mathcal{E}$ and of $\pi_{xy}^{-1}(p)\cap\overline{\mathcal{E}}$ is the same. We use the \verb?RAG? package to get one point in each connected component and \verb?RS? to solve the corresponding zero dimensional real systems. This allows us to prove that 
$$
\pi_{xy}^{-1}(\mathbb{R}^2\setminus (V_1^{xy}\cup V_2^{xy}))\cap\mathcal{E}=\pi_{xy}^{-1}(\mathbb{R}^2\setminus (V_1^{xy}\cup V_2^{xy}))\cap\overline{\mathcal{E}}
$$

\begin{figure}
  \begin{tabular}{@{}c@{ }c}
  \includegraphics[angle=-90,width=0.5\textwidth]{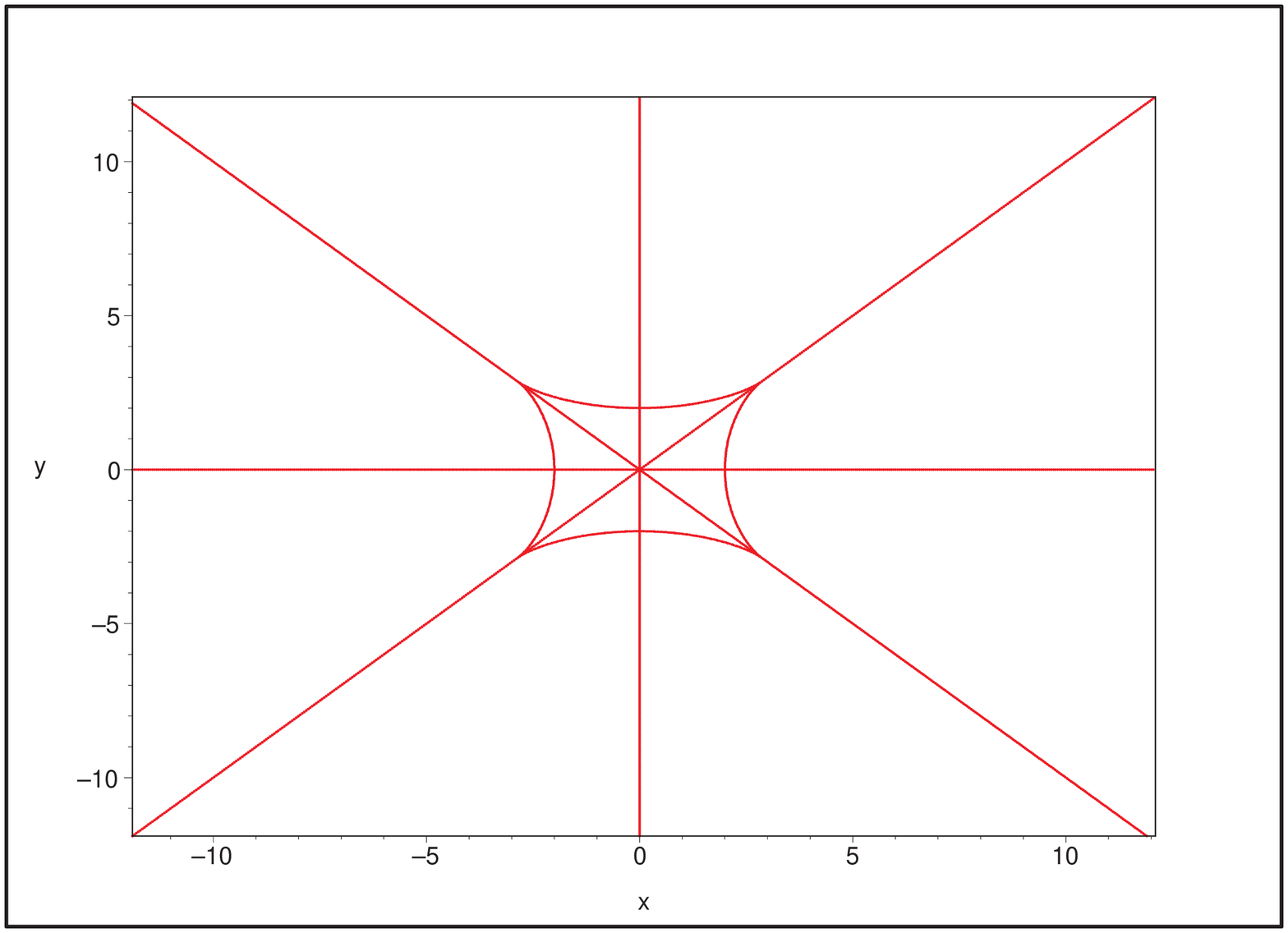} &
  \includegraphics[angle=-90,width=0.5\textwidth]{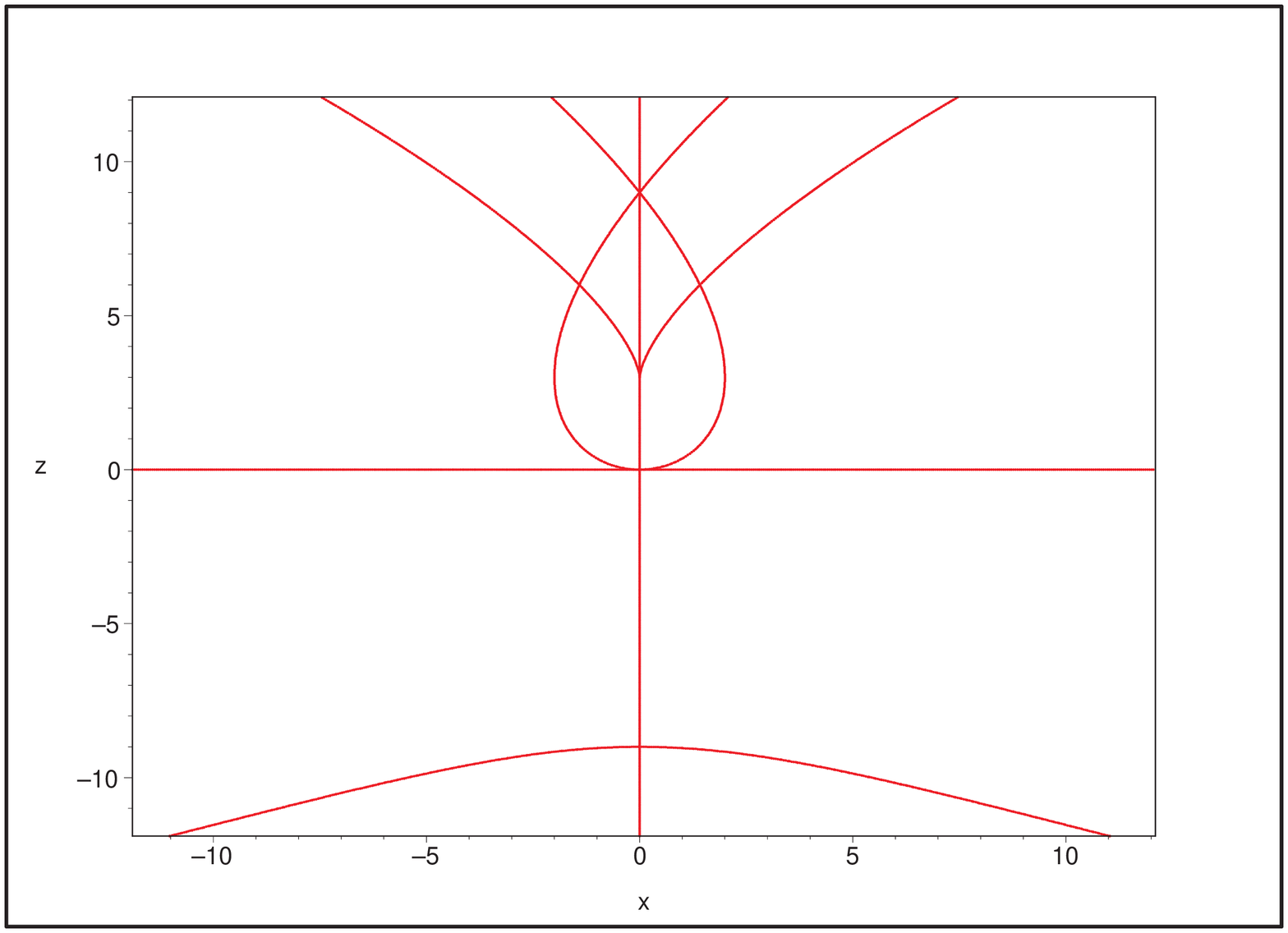}\\\\
  $V_1^{xy}\cup V_2^{xy}$ & $V_1^{xz}\cup V_2^{xz}$
  \end{tabular}
  \begin{center}
  \includegraphics[angle=-90,width=0.5\textwidth]{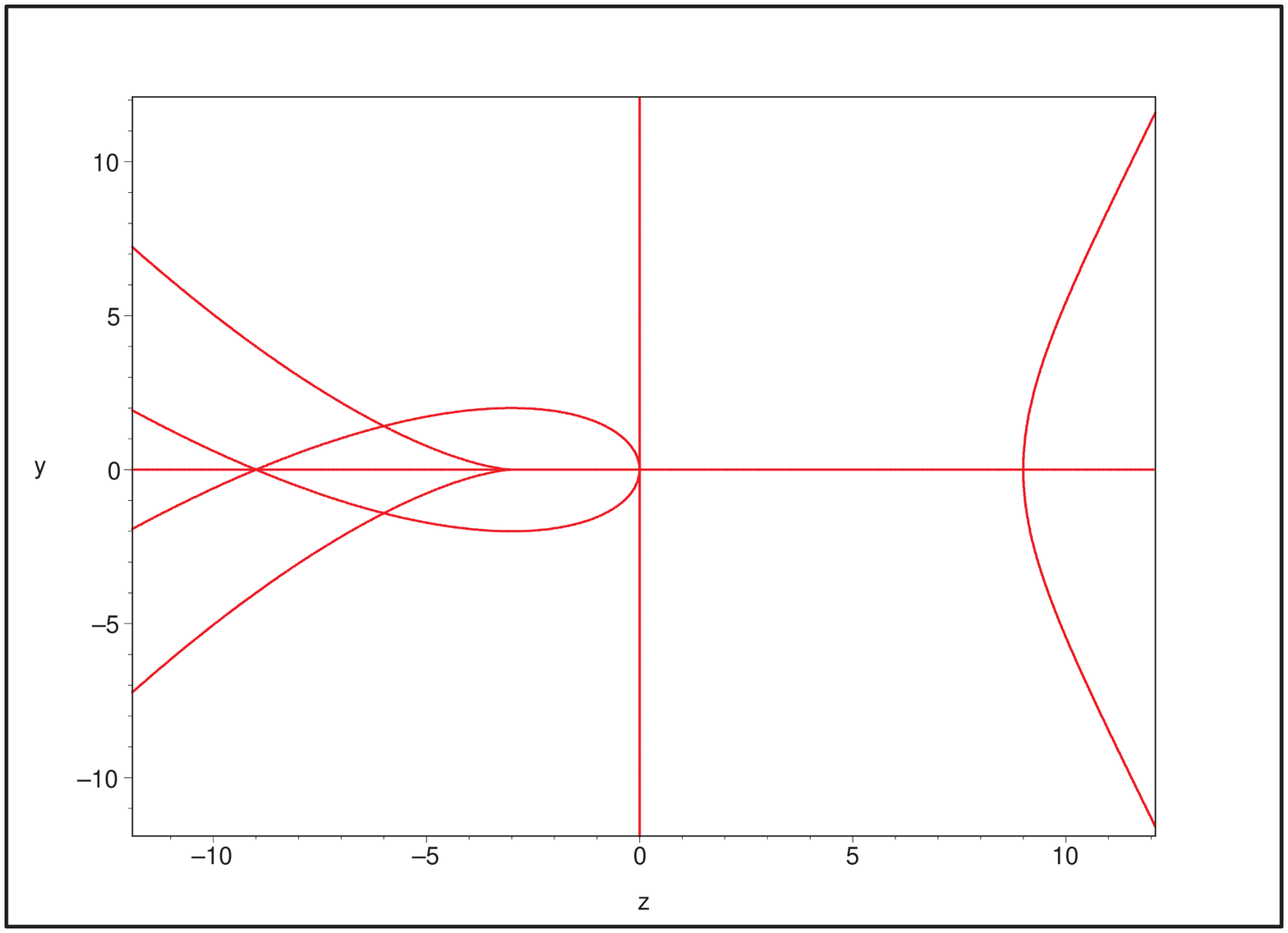}\bigskip\\
  $V_1^{yz}\cup V_2^{yz}$
  \end{center}

  \caption{The discriminant varieties for the three possible sets of parameters}
  \label{discrvar}
\end{figure}

In order to get more information, we repeat this process using respectively the discriminant varieties on the parameter set $\{x,z\}$ and $\{y,z\}$. This leads to the following computations:
$$\begin{array}{l@{\ }c@{\ ,\ }c@{{\ },{\ }}c@{{\ },{\ }}c@{{\ })}}
  V_1^{xz}:=\mathbf{DV}(&[x-x(u,v),y-y(u,v),z-z(u,v)]&[\ ]&[x,z]&[y,u,v]\\
  V_2^{xz}:=\mathbf{DV}(&[p(x,y,z)]&[\ ]&[x,z]&[y]\\
\end{array}$$
and
$$\begin{array}{l@{\ }c@{\ ,\ }c@{{\ },{\ }}c@{{\ },{\ }}c@{{\ })}}
  V_1^{yz}:=\mathbf{DV}(&[x-x(u,v),y-y(u,v),z-z(u,v)]&[\ ]&[y,z]&[x,u,v]\\
  V_2^{yz}:=\mathbf{DV}(&[p(x,y,z)]&[\ ]&[y,z]&[x]\\
\end{array}
$$
The result is shown on Figure \ref{discrvar}.

Then we compute as above one point in each connected component of the complementary, and this allows us to prove that:
$$
\pi_{xz}^{-1}(\mathbb{R}^2\setminus (V_1^{xz}\cup V_2^{xz}))\cap\mathcal{E}=\pi_{xz}^{-1}(\mathbb{R}^2\setminus (V_1^{xz}\cup V_2^{xz}))\cap\overline{\mathcal{E}}
$$
and
$$
\pi_{yz}^{-1}(\mathbb{R}^2\setminus (V_1^{yz}\cup V_2^{yz}))\cap\mathcal{E}=\pi_{yz}^{-1}(\mathbb{R}^2\setminus (V_1^{yz}\cup V_2^{yz}))\cap\overline{\mathcal{E}}
$$

Using the following notations:
$$\begin{array}{c}
  V^{xy}:=\pi_{yz}^{-1}(V_1^{xy}\cup V_2^{xy})\\
  V^{xz}:=\pi_{xz}^{-1}(V_1^{xz}\cup V_2^{xz})\\
  V^{yz}:=\pi_{yz}^{-1}(V_1^{yz}\cup V_2^{yz})
\end{array}$$

it remains us to check what happens above each component of
$$V^{xy}\cap V^{xz}\cap V^{yz}$$
An idea is to set apart the linear components from the others. We introduce $$V_L:=\mathbf{V}(x+y)\cup \mathbf{V}(x-y)\cup \mathbf{V}(x)\cup \mathbf{V}(y)\cup \mathbf{V}(z)$$
and denote respectively $V^{xy}\setminus V_L$,$V^{xz}\setminus V_L$ and $V^{yz}\setminus V_L$ by $\widetilde{V^{xy}}$, $\widetilde{V^{xz}}$ and $\widetilde{V^{yz}}$. Using the RAGlib, we verify that $$\widetilde{V^{xy}}\cap\widetilde{V^{xz}}\cap\widetilde{V^{yz}}$$ has actually no real points.

It remains us to check what happens on each of the 5 linear components of $V_L$. The intersection of $\mathcal{E}_g$ or $\overline{\mathcal{E}}$ with a linear component $P$ may be seen as a linear substitution of a variable. This operation produces 5 pairs of varieties of dimension 2 (Table \ref{table}). To check their equality, we use the same strategy as above and compute the 5 discriminant varieties with 1 parameter,3 unknowns of $K_1,\ldots,K_5$, respectively $V_{K_1},\ldots,V_{K_5}$, and the 5 discriminant varieties with 1 parameter,1 unknown of $L_1,\ldots,L_5$, respectively $V_{L_1},\ldots,V_{L_5}$. We check that $K_i=L_i$ for each point by connected component of the complementary of $V_{K_i}\cup V_{L_i}$, in less than 1 second. And at last we intersect again the varieties with their discriminant varieties, which reduces the problem to compare 5 pairs of zero dimensional systems. Thus we check that the equality holds for the finitely many points considered. Finally this allows us to conclude that $\mathcal{E}=\overline{\mathcal{E}}$.

\begin{sidewaystable}
  $$\begin{array}{|c|c|c|c|c|c|c|}
    \cline{2-6}
    \multicolumn{1}{c}{} &\multicolumn{5}{|c|}{\mathcal{E}_g}\\
    \hline
    W &\mathbf{V}(x) & \mathbf{V}(y) & \mathbf{V}(z) & \mathbf{V}(y+x) & \mathbf{V}(y-x)\\
    \hline
    \mathcal{E}_g\cap W& K_1 & K_2 & K_3 & K_4 & K_5\\
    \hline
    System & \left\{\begin{array}{@{}l@{}}0-x(u,v)\\y-y(u,v)\\z-z(u,v)\end{array}\right.
      & \left\{\begin{array}{@{}l@{}}x-x(u,v)\\0-y(u,v)\\z-z(u,v)\end{array}\right.
      & \left\{\begin{array}{@{}l@{}}x-x(u,v)\\y-y(u,v)\\0-z(u,v)\end{array}\right.
      & \left\{\begin{array}{@{}l@{}}x-x(u,v)\\-x-y(u,v)\\z-z(u,v)\end{array}\right.
	& \left\{\begin{array}{@{}l@{}}x-x(u,v)\\x-y(u,v)\\z-z(u,v)\end{array}\right.
      \\
    \hline
    Parameter & z & z & x & x & x\\
    \hline
    Unknowns & y,u,v & x,u,v & y,u,v & z,u,v & z,u,v\\
    \hline
    \parbox{2cm}{\centering $Minimal$\\$Discriminant$\\$Variety$} &
%   \multicolumn{2}{|c|}{
\begin{array}{@{}c@{}}
      V_{K_1}=\\
      \mathbf{V}(z)\cup\mathbf{V}(z-3)\\
      \cup\mathbf{V}(z-9)\\
    \end{array}
    &
    \begin{array}{@{}c@{}}
      V_{K_2}=\\
      \mathbf{V}(z)\cup\mathbf{V}(z+3)\\
      \cup\mathbf{V}(z+9)\\
    \end{array}

%    }
    & 
    \begin{array}{@{}c@{}}
      V_{K_3}=\\
      \mathbf{V}(x)\cup\mathbf{V}(x^2+2)\\
      \\
    \end{array}

    &
    \multicolumn{2}{|c|}{
    \begin{array}{@{}c@{}}
      V_{K_4}=V_{K_5}=\\
      \mathbf{V}(x+4)\cup\mathbf{V}(x-4)\cup\mathbf{V}(x^2-8)\\
      \cup\mathbf{V}(x^2+2)\\
    \end{array}
    }
    \\
    \hline
    \multicolumn{6}{c}{}\\
    \multicolumn{6}{c}{}\\
    \multicolumn{6}{c}{}\\
    \multicolumn{6}{c}{}\\
    \cline{2-6}
    \multicolumn{1}{c}{} &\multicolumn{5}{|c|}{\overline{\mathcal{E}}}\\
    \hline
    W &\mathbf{V}(x) & \mathbf{V}(y) & \mathbf{V}(z) & \mathbf{V}(y+x) & \mathbf{V}(y-x)\\
    \hline
    \overline{\mathcal{E}}\cap W& L_1 & L_2 & L_3 & L_4 & L_5\\
    \hline
    \begin{array}{c}System\\(sqfr =\\squarefree)\end{array} & sqfr(p(0,y,z))
      & sqfr(p(x,0,z))
      & sqfr(p(x,y,0))
      & sqfr(p(x,-x,z))
      & sqfr(p(x,x,z))
      \\
    \hline
    Parameter & z & z & x & x & x\\
    \hline
    Unknown & y & x & y & z & z\\
    \hline
    \parbox{2cm}{\centering $Minimal$\\$Discriminant$\\$Variety$} &
%    \multicolumn{2}{|c|}{
    \begin{array}{@{}c@{}}
      V_{L_1}=\mathbf{V}(z+9)\\
      \cup\mathbf{V}(z)\cup\mathbf{V}(z-3)\\
      \cup\mathbf{V}(z-9)\\
    \end{array}
    &
    \begin{array}{@{}c@{}}
      V_{L_2}=\mathbf{V}(z-9)\\
      \cup\mathbf{V}(z)\cup\mathbf{V}(z-3)\\
      \cup\mathbf{V}(z+9)\\
    \end{array}

%   }
    & 
    \begin{array}{@{}c@{}}
      V_{L_3}=\\
      \mathbf{V}(x)\\
      \\
    \end{array}

    &
    \multicolumn{2}{|c|}{
    \begin{array}{@{}c@{}}
      V_{L_4}=V_{L_5}=\\
      \mathbf{V}(x+4)\cup\mathbf{V}(x-4)\cup\mathbf{V}(x^2-8)\\
      \cup\mathbf{V}(x)\\
    \end{array}
    }
    \\
    \hline
  \end{array}$$
  \caption{Discriminant varieties of the sub varieties}
  \label{table}
\end{sidewaystable}

%$$\begin{array}{ll}
%  V_1^{xy}\cup V_2^{xy}= &\mathbf{V}(x)\\
%&\cup \mathbf{V}(y)\\
%&\cup \mathbf{V}(x+y)\\
%&\cup \mathbf{V}(x-y)\\
%&\cup\mathbf{V}(y^6+60y^4+768y^2-4096+3x^2y^4-312x^2y^2+768x^2+3x^4y^2+60x^4+x^6)\\
%  &\cup\mathbf{V}(x^6+48x^4+3x^4y^2-336x^2y^2+3x^2y^4+768x^2+4096+768y^2+48y^4+y^6)\\
%\end{array}
%$$

\section{Conclusion}
We provided a \emph{deterministic} single exponential bit-com\-plexity bound for the computation of the minimal discriminant variety of a \emph{generically simple} parametric system. Note that the complexity of our algorithm relies on the elimination problem's complexity. Thus in a probabilistic bounded Turing machine, the work of \cite{MR1624458} for example leads to a polynomial complexity bound in the size of the output. Or if we are only interested in the real solutions, then the use of the single block elimination routine of \cite{BasPolRoy96,Basu99a} improves directly the deterministic complexity bound of our method.

The reduction presented in this article is easy to implement in conjunction with a software performing elimination, as those used in \cite{Faugere:2002:NEA,MR2093185}, \cite{Cox-Little-O_Shea/98} or \cite{GiustiSchost:1999} for example.

It would be worth studying the complexity of the computation of the \emph{minimal} discriminant variety when we have more equations than unknowns.

\pagebreak


\begin{thebibliography}{10}

\bibitem{SALSA}
{SALSA}: http://fgbrs.lip6.fr/salsa/software.

\bibitem{conf/issac/AnaiHY05}
{\sc Anai, H., Hara, S., and Yokoyama, K.}
\newblock Sum of roots with positive real parts.
\newblock In {\em ISSAC'05}, pp.~21--28.

\bibitem{Basu99a}
{\sc Basu}.
\newblock New results on quantifier elimination over real closed fields and
  applications to constraint databases.
\newblock {\em JACM: Journal of the ACM 46\/} (1999).

\bibitem{BasPolRoy96}
{\sc Basu, Pollack, and Roy}.
\newblock On the combinatorial and algebraic complexity of quantifier
  elimination.
\newblock {\em JACM: Journal of the ACM 43\/} (1996).

\bibitem{MR1213453}
{\sc Becker, T., and Weispfenning, V.}
\newblock {\em Gr\"obner bases}, vol.~141 of {\em Graduate Texts in
  Mathematics}.
\newblock Springer-Verlag, New York, 1993.

\bibitem{oai:CiteSeerPSU:552837}
{\sc Bernasconi, A., Mayr, E.~W., Raab, M., and Mnuk, M.}
\newblock Computing the dimension of a polynomial ideal, May~06 2002.

\bibitem{conf/issac/BrownM05}
{\sc Brown, C.~W., and McCallum, S.}
\newblock On using bi-equational constraints in {CAD} construction.
\newblock In {\em ISSAC\/} (2005), pp.~76--83.

\bibitem{MR1653279}
{\sc Brownawell, W.~D.}
\newblock A pure power product version of the {H}ilbert {N}ullstellensatz.
\newblock {\em Michigan Math. J. 45}, 3 (1998), 581--597.

\bibitem{Buchberger:1976:TBR}
{\sc Buchberger, B.}
\newblock A theoretical basis for the reduction of polynomials to canonical
  forms.
\newblock {\em j-SIGSAM 10}, 3 (Aug. 1976), 19--29.

\bibitem{Collins75}
{\sc Collins, G.~E.}
\newblock {\em Quantifier Elimination for Real Closed Fields by Cylindrical
  Algebraic Decomposition}.
\newblock Springer Verlag, 1975.

\bibitem{conf/adg/CorvezR02}
{\sc Corvez, S., and Rouillier, F.}
\newblock Using computer algebra tools to classify serial manipulators.
\newblock In {\em Automated Deduction in Geometry\/} (2002), pp.~31--43.

\bibitem{Cox92}
{\sc Cox, D., Little, J., and O'Shea, D.}
\newblock {\em Ideals, Varieties, and Algorithms}.
\newblock Undergraduate Texts in Mathematics. Springer Verlag, 1992.

\bibitem{Cox-Little-O_Shea/98}
{\sc Cox, D.~A., Little, J.~B., and O'Shea, D.~B.}
\newblock {\em Using algebraic geometry}, vol.~185 of {\em Graduate Texts in
  Mathematics}.
\newblock Springer-Verlag, 1998.

\bibitem{conf/adg/Dolzmann98}
{\sc Dolzmann, A.}
\newblock Solving geometric problems with real quantifier elimination.
\newblock In {\em Automated Deduction in Geometry\/} (1998), vol.~1669 of {\em
  Lecture Notes in Computer Science}, pp.~14--29.

\bibitem{Faugere:2002:NEA}
{\sc Faug{\`e}re, J.~C.}
\newblock A new efficient algorithm for computing {Gr{\"o}bner} bases without
  reduction to zero {$(F_5)$}.
\newblock In {\em {ISSAC} 2002\/} (2002), pp.~75--83.

\bibitem{MR2093185}
{\sc Faug{\`e}re, J.-C., and Joux, A.}
\newblock Algebraic cryptanalysis of hidden field equation ({HFE})
  cryptosystems using {G}r\"obner bases.
\newblock In {\em Advances in cryptology---CRYPTO 2003}, vol.~2729 of {\em
  Lecture Notes in Comput. Sci.} Springer, Berlin, 2003, pp.~44--60.

\bibitem{FRPM:ijs:2006}
{\sc Fotiou, I.~A., Rostalski, P., Parrilo, P.~A., and Morari, M.}
\newblock Parametric optimization and optimal control using algebraic geometry
  methods.
\newblock {\em International Journal of Control (to appear)\/}.

\bibitem{MR1644323}
{\sc Fulton, W.}
\newblock {\em Intersection theory}, second~ed., vol.~2.
\newblock Springer-Verlag, Berlin, 1998.

\bibitem{conf/eurosam/Giusti84}
{\sc Giusti, M.}
\newblock Some effectivity problems in polynomial ideal theory.
\newblock In {\em EUROSAM 84}, vol.~174 of {\em Lecture Notes in Comput. Sci.}
  1984, pp.~159--171.

\bibitem{Giusti-Heintz/93}
{\sc Giusti, M., and Heintz, J.}
\newblock La d{\'e}termination des points isol{\'e}s et de la dimension d'une
  vari{\'e}t{\'e} alg{\'e}brique peut se faire en temps polynomial.
\newblock In {\em Computational algebraic geometry and commutative algebra},
  vol.~34. 1993, pp.~216--256.

\bibitem{GiustiSchost:1999}
{\sc Giusti, M., and Schost, {\'E}.}
\newblock Solving some overdetermined polynomial systems.
\newblock In {\em ISSAC'99}.

\bibitem{conf/issac/GrigorievV00}
{\sc Grigoriev, D., and Vorobjov, N.}
\newblock Bounds on numers of vectors of multiplicities for polynomials which
  are easy to compute.
\newblock In {\em ISSAC'00}, pp.~137--146.

\bibitem{TCS::Heintz1983}
{\sc Heintz, J.}
\newblock Definability and fast quantifier elimination in algebraically closed
  fields.
\newblock {\em Theoretical Computer Science 24}, 3 (Aug. 1983), 239--277.

\bibitem{MR1414452}
{\sc Krick, T., and Pardo, L.~M.}
\newblock A computational method for {D}iophantine approximation.
\newblock In {\em Algorithms in algebraic geometry and applications}, vol.~143
  of {\em Progr. Math.} 1996, pp.~193--253.

\bibitem{Lazard/77}
{\sc Lazard, D.}
\newblock Alg{\`e}bre lin{\'e}aire sur ${K}[{X}_1,\ldots,{X}_n]$ et
  {\'e}limination.
\newblock {\em Bull.~Soc.~Math.~France 105\/} (1977), 165--190.

\bibitem{rouillier/lazard}
{\sc Lazard, D., and Rouillier, F.}
\newblock Solving parametric polynomial systems.
\newblock Tech. rep., INRIA, 2004.
\newblock Accepted for publication in Journal of Symbolic Computation.

\bibitem{Matera97}
{\sc Matera, G., and Turull~Torres, J.~M.}
\newblock The space complexity of elimination theory: upper bounds.
\newblock In {\em Foundations of computational mathematics}. 1997,
  pp.~267--276.

\bibitem{MR1624458}
{\sc Puddu, S., and Sabia, J.}
\newblock An effective algorithm for quantifier elimination over algebraically
  closed fields using straight line programs.
\newblock {\em J. Pure Appl. Algebra 129}, 2 (1998), 173--200.

\bibitem{MR1512592}
{\sc Rabinowitsch, J.~L.}
\newblock Zum {H}ilbertschen {N}ullstellensatz.
\newblock {\em Math. Ann. 102}, 1 (1930), 520.

\bibitem{MR1959170}
{\sc Schost, {\'E}.}
\newblock Computing parametric geometric resolutions.
\newblock {\em Appl. Algebra Engrg. Comm. Comput. 13}, 5 (2003), 349--393.

\bibitem{MR2161992}
{\sc Sommese, A.~J., Verschelde, J., and Wampler, C.~W.}
\newblock Introduction to numerical algebraic geometry.
\newblock In {\em Solving polynomial equations}, vol.~14 of {\em Algorithms
  Comput. Math.} 2005, pp.~301--335.

\bibitem{conf/issac/Verschelde04}
{\sc Verschelde, J.}
\newblock Numerical algebraic geometry and symbolic computation.
\newblock In {\em ISSAC\/} (2004), p.~3.

\bibitem{MR1826878}
{\sc Wang, D.}
\newblock {\em Elimination methods}.
\newblock Springer-Verlag, Vienna, 2001.

\bibitem{Weispfenning92}
{\sc Weispfenning}.
\newblock Comprehensive grobner bases.
\newblock {\em Journal of Symbolic Computation 14\/} (1992).

\bibitem{Yang:2005:OPM}
{\sc Yang, L., and Zeng, Z.}
\newblock An open problem on metric invariants of tetrahedra.
\newblock In {\em {ISSAC} '05}, pp.~362--364.

\end{thebibliography}
\end{document}